\documentclass[prl,nofootinbib,twocolumn]{revtex4}
\usepackage{graphicx}
\usepackage{bm}
\usepackage{float}
\usepackage{subfigure}
\usepackage{amsmath}
\usepackage{amsfonts}

\def\be{\begin{equation}}
\def\ee{\end{equation}}
\def\bed{\begin{description}}
\def\eed{\end{description}}

\def\bea{\begin{eqnarray}}
\def\eea{\end{eqnarray}}

\def\ba{\begin{array}}
\def\ea{\end{array}}

\def\half{\frac{1}{2}\,}

\def\quart{\frac{1}{4}\,}

\def\su{$SU(2)$}
\def\u1{$U(1)$}
\def\suu1{$SU(2)\times U(1)$}

\def\du{D_\mu}

\def\Du{D^\mu}

\def\bun{B_{\mu \nu}}
\def\Bun{B^{\mu \nu}}
\def\wun{\bm{W}_{\mu \nu}}
\def\Wun{\bm{W}^{\mu \nu}}
\def\bu{B_\mu}

\def\wu{\bm{W}_\mu}

\def\nn{\nonumber}

\begin{document}

\title{Helical Magnetic Fields from Sphaleron Decay and Baryogenesis}

\author{Craig J. Copi$^1$, Francesc Ferrer$^1$, Tanmay Vachaspati$^{1,2}$
and Ana Ach\'ucarro$^{3,4}$}
\affiliation{$^1$CERCA, Department of Physics, Case Western Reserve University,
10900 Euclid Avenue, Cleveland, OH 44106-7079, \\
{$^2$Institute for Advanced Study, Princeton, NJ 08540} \\
{$^3$Instituut-Lorentz for Theoretical Physics, Leiden, The Netherlands,} \\
{$^4$Department of Theoretical Physics, 
The University of the Basque Country UPV-EHU, 38940 Bilbao, Spain}}

\begin{abstract}
\noindent
Many models of baryogenesis rely on anomalous particle physics processes 
to give baryon number violation. By numerically evolving the electroweak 
equations on a lattice, we show that baryogenesis in these models creates 
helical cosmic magnetic fields. After a transitory period, electroweak 
dynamics is found to conserve the Chern-Simons number and the total 
electromagnetic helicity. We argue that baryogenesis could lead to
magnetic fields of nano-Gauss strength today on astrophysical
length scales. In addition to being astrophysically relevant, 
such helical magnetic fields can provide an independent probe of 
baryogenesis and CP violation in particle physics. 
\end{abstract}

\maketitle

Cosmic magnetic fields can arise from a number of early universe
processes ({\it e.g.} \cite{Turner:1987bw,Vachaspati:1991nm,
Ratra:1991bn,Vachaspati:1994xc,Cornwall:1997ms,Vachaspati:2001nb}).
The detection of such magnetic fields would be an important 
step in our understanding of structure evolution, and the magnetic
field distribution could be used as a probe of early universe 
cosmology. Most magnetic field generation mechanisms discussed 
so far, produce non-helical fields, and thus it is of particular 
interest that processes like electroweak baryogenesis imply magnetic 
fields that are helical \cite{Cornwall:1997ms,Vachaspati:2001nb}.
If helical primordial magnetic fields are observed, they would
offer confirmation of cosmological baryogenesis, and the magnetic
field properties could be turned into a detailed probe of particle
physics and cosmology at the epoch of baryogenesis. In particular,
the CP violation that leads to a universe filled with matter and 
no antimatter would be probed by cosmological observations.

These considerations have prompted us to study the production
of magnetic fields during the decay of sphalerons \cite{Manton:1983nd,
Klinkhamer:1984di}, a process that is key to electroweak baryogenesis. 
While electroweak baryogenesis is not successful in the minimal 
standard model, our study also applies to any extension of the 
standard model in which baryon number violation is mediated by 
sphaleron-like processes involving changes in the winding 
(Chern-Simons number) of vacuum non-Abelian gauge field 
configurations. 

To understand the connection between sphalerons and helical
magnetic fields, it is simpler to think of ``deformed sphalerons''
where the gauge field configuration resembles that of electroweak
strings \cite{Vachaspati:1992fi}. The sphaleron can then be
interpreted as linked loops of electroweak Z-string or a 
confined electroweak monopole-antimonopole pair 
\cite{Vachaspati:1994ng,Hindmarsh:1993aw,Garriga:1994wb,
Achucarro:1999it}. 
The linked loops of Z magnetic flux can then decay into linked 
electromagnetic flux as described in \cite{Vachaspati:2001nb} 
and thus the resulting electromagnetic field carries magnetic 
helicity. If we think of the sphaleron in terms of the confined 
magnetic monopole pair, the string that confines them is twisted 
and this also leads to magnetic helicity. 
In \cite{Cornwall:1997ms,Vachaspati:2001nb} 
these considerations indicated a remarkably simple relationship
between the cosmic magnetic helicity density and the cosmic baryon 
number density
\begin{equation}
h = \frac{1}{V} \int_V d^3 x ~ {\bf A}\cdot {\bm \nabla}\times {\bf A}
  ~ \sim ~ - \frac{n_b}{\alpha}
\label{hnb}
\end{equation}
where we consider some large spatial volume $V$, ${\bf A}$ is 
the electromagnetic vector potential, $n_b$ is the average baryon 
number density and $\alpha =1/137$ is the fine structure constant.

Our goal is to examine the heuristic arguments in 
\cite{Cornwall:1997ms,Vachaspati:2001nb} by explicitly studying 
the decay of an electroweak sphaleron. (Recently, along similar
lines, magnetic fields produced during preheating at the electroweak 
scale
were studied 
in Ref.~\cite{DiazGil:2007dy}.)
We will indeed find that helical magnetic fields are generated 
during sphaleron decay and the relation in Eq.~(\ref{hnb}) holds 
at the order of magnitude level. Our numerical results also
show, somewhat unexpectedly, that the Chern-Simons number and 
the electromagnetic helicity are conserved after a transitory 
initial period (also see \cite{Jackiw:1999bd}). We also reconsider 
the net magnetic field generated during baryogenesis and find 
that the field strength is likely to be much larger than has 
been previously estimated. 

We work with the bosonic sector of the electroweak Lagrangian 
\bea
{\cal L}&=&\left(\du \Phi\right)^\dagger \Du \Phi-\quart \bun \Bun -
\quart \wun \cdot \Wun \nn \\
&-& \lambda \left(|\Phi|^2-v^2 \right)^2
\label{eq:lagrangian}
\eea
where
\be
\du \Phi = \left(\partial_\mu - i \frac{g'}{2} \bu -
                  i \frac{g}{2} \bm{\tau} \cdot \wu 
                      \right) \Phi,
\label{eq:dphi}
\ee
The $SU(2)$ generators, $\tau^a$, are normalized by
${\rm Tr}(\tau^a \tau^b)=2\delta^{ab}$. The \u1\ (hypercharge) 
field strength is
\be
\bun = \partial_\mu B_\nu - \partial_\nu \bu,
\label{eq:bmunu}
\ee
and the \su\ field strength is defined as
\be
W_{\mu \nu}^a  = \partial_\mu W_\nu^a - \partial_\nu W_\mu^a 
+ g \epsilon^{b c a}  W_\mu^b W_\nu^c
\label{eq:compwmunu}
\ee

To study the magnetic field produced by a decaying sphaleron, we 
numerically set up a configuration like the electroweak sphaleron 
solution \cite{Manton:1983nd,Klinkhamer:1984di, Kunz:1992uh},
{\it i.e.} a perturbed sphaleron, since the exact solution, even
though it is unstable, will take a long time to decay numerically.
\be
\Phi = v h(\xi) {\bf G}_\Theta 
\begin{pmatrix}
i x + y \\
-i z
\end{pmatrix}
\ee

\bea
W_i^a {\bf \tau}^a = -\frac{2 f(\xi)}{g r^2} \epsilon_{i c b} x_b {\bf G}_\Theta  
 {\bf \tau}^c {\bf G}_\Theta^\dagger 
\label{gwbarmua}
\eea

\be
B_i = g' v^2 p(\xi) \left(-y, x, 0\right)
\label{gpbarymu}
\ee

where
\begin{displaymath}
{\bf G}_\Theta \left(\vec{x}\right)=\exp{\left[i \Theta(r) 
{\bm \tau} \cdot \hat{\bf x}/2 \right]}
\end{displaymath}
is an $SU(2)$ gauge transformation, with $\Theta(r)=\pi \left(1-\exp(-r)
\right)$, ensuring that the gauge fields fall off fast enough away from
the origin.
The functions $f(\xi)$, $h(\xi)$ and $p(\xi)$ are profile functions and
are taken to depend on the radial coordinate $\xi=g v r/\sqrt{2}$ alone.

In the true solution, the profile functions also depend on 
the angular coordinates and have to be determined numerically
\cite{Kunz:1992uh}. Since we want to start with a perturbed 
sphaleron we do not need the precise forms of $f$ and $h$, 
but we do need to meet the asymptotic properties satisfied by 
$f$ and $h$: $f, ~h \to 0$ as $r \to 0$, and 
$f, ~h \to 1$ as $r \to \infty$. 
Following Klinkhamer and Manton \cite{Klinkhamer:1984di},
we choose for the profile functions their {\it Ansatz}, labeled {\it a},
with the length scales given by $\Xi = 3.79$ and $\Omega = 1.90$. 
We take for the initial $U(1)$ gauge field, Eq.~(\ref{gpbarymu}), the small
$g'$ approximation~\cite{Klinkhamer:1984di}. The profile function $p$ can
be found, in terms of $f$ and $h$, by solving the $U(1)$ equation of motion
in the $SU(2)$ background.

The decay is studied by evolving the $SU(2)\times U(1)$ electroweak 
field equations using the standard Wilsonian approach for lattice gauge
fields~\cite{Ambjorn:1990pu,Moore:1996wn,Tranberg:2003gi,GarciaBellido:2003wd,
Graham:2006vy}.
The temporal gauge, $W_0^a=B_0=0$, allows a simple identification of the
canonical momentum, and we use it throughout in this work.
We adopt Graham's implementation of the lattice equations 
\cite{Graham:2006vy}
  \begin{eqnarray}
    \phi (t+) &=& 2\phi (t) - \phi (t-) +(\Delta t)^2 \biggl[
 2 \lambda \left(v^2 -|\phi^p|^2\right)\phi^p 
 \nonumber \\
  && + \sum_{i=x,y,z}{ \frac{U_i^p \phi^{p+i} -2 \phi^p +
          U_{-i}^p \phi^{p-i}}{\Delta x^2}}
~ \biggr  ]
  \end{eqnarray}
where the gauge fields are on the links and are defined in terms
of the $U_j^p$ matrices by
\be
U_j^p = {\rm e}^{i g' B_j^p \Delta x/2}
{\rm e}^{i g \bm{W}_j^p \cdot \bm{\tau}\Delta x/2}.
\label{eq:link}
\ee
Here $p$ labels the point in the lattice and $j$ the link
emanating from point $p$. 
The $U$ matrix along a timelike link, $j=t$, is equal to the identity
matrix in the temporal gauge.
For the link from site $p$ in the negative $j$ direction, we take:
\be
U_{-j}^p = \left(U_j^{p-j} \right)^\dagger =
{\rm e}^{-i g' B_j^{p-j} \Delta x/2}
{\rm e}^{-i g \bm{W}_j^{p-j} \cdot \bm{\tau} \Delta x/2}.
\ee
The evolution of the gauge fields is given by
  \begin{eqnarray}
    U_j^p (t+) &=& \exp \biggl [ \log{U^p_j (t) U^p_j{}^\dagger(t_-)}
     \nonumber \\
    && -\frac{\Delta t^2}{\Delta x^2} \sum_{j' \neq j} {\left(\log {
          U_{\Box j j'}^p} + \log { U_{\Box j -j'}^p}\right)} 
     \nonumber \\
    &&
      + \half i
    \Delta t^2 \Delta x \left( 
      g' J_j^p + g \bm{J}_j^p\cdot \bm{\tau} \right) \biggr ] U_j^p
  \end{eqnarray}
Here, we define $U_{\Box j j'}^p \equiv U^p_j U^{p+j}_{j'} U^{p+j+j'}_{-j}
U^{p+j'}_{-j'}$ for the plaquette and $J_j^p\equiv -g'{\rm Im}\left( \phi^p
  {}^\dagger U_j^p \phi^{p+j}\right)/\Delta x$ and $\bm{J}_j^p\equiv -g'{\rm
  Im}\left( \phi^p {}^\dagger \bm{\tau} U_j^p \phi^{p+j}\right)/\Delta x$
for the gauge currents.  For further details, the reader is referred to
\cite{Graham:2006vy}.  In addition to these evolution equations we
have implemented absorbing boundary conditions as described
in~\cite{Olum:1999sg} but extended to non-Abelian fields.

The evolution of Chern-Simons number is of particular interest
to us since it is correlated with changes in the baryon number. It is
given by
\begin{eqnarray}
N_{CS} (t) &=& \frac{N_F}{32\pi^2} \epsilon^{ijk} \int d^3 x  \biggl [ 
- {g'}^2 B_{ij} B_k  
            \nonumber \\
&+& g^2 \left ( W_{ij}^a W_k^a - \frac{g}{3} 
                      \epsilon_{abc} W_i^a W_j^b W_k^c \right )
\biggr ]
\label{NCSformula}
\end{eqnarray}
where $N_F =3$ is the number of families and there is no surface
term because our fields vanish at infinity.



In the electroweak model, the electromagnetic gauge field $A_\mu$ 
is defined in terms of the electroweak $W^\mu$ and $B^\mu$ gauge 
fields
\begin{equation}
A_\mu = \sin\theta_w n^a W_\mu^a+ \cos\theta_w B_\mu 
\label{amu}
\end{equation}
with $n^a = - \Phi^\dag {\bf \tau}^a \Phi / \Phi^\dag \Phi$.
However, there is a choice of definitions for the electromagnetic
field strength. For example, 
\begin{equation}
A_{\mu \nu} = \sin\theta _w ~n^a W_{\mu \nu} ^a +
               \cos\theta _w ~B_{\mu \nu} 
\label{Amunudefn}
\end{equation}
is the natural definition to calculate
the energy density in the massless electromagnetic field. 
Yet this definition is not simply the curl of the gauge 
field in Eq.~(\ref{amu}) because derivatives of $A_\mu$ also 
involve derivatives of $n^a$. Even at late times, when $D_\mu\Phi$
becomes small, such derivatives will in general contribute. The 
only clean resolution of this issue that we have found is to define 
the electromagnetic gauge field in unitary gauge and then define the 
electromagnetic field strength as the curl of the gauge potential 
as in the usual 
Maxwell theory. At early times, when $D_\mu \Phi$ is significant, 
${\bm \nabla}\times {\bf A}$ does not coincide with $A_{ij}$ in 
Eq.~(\ref{Amunudefn}), but they do coincide at late times. Then 
the helicity tells us something about the topology of the very same 
magnetic field lines that carry energy density.

To go into unitary gauge, we calculate $A_\mu$ at every time
step only after applying the SU(2) gauge rotation
\begin{equation}
g_2 = \frac{1}{\sqrt{\Phi^\dag \Phi}}
       \begin{pmatrix}
                \Phi_2   & -\Phi_1 \\
                \Phi_1^* & \Phi_2^* 
       \end{pmatrix}
\end{equation}
where $\Phi_1$ and $\Phi_2$ are the upper and lower components
of $\Phi$. It is easily verified that 
$g_2 \Phi = \sqrt{\Phi^\dag \Phi} (0,1)^T$, ${\rm det}g_2 =1$ 
and $g_2^\dag = g_2^{-1}$. After the rotation, 
$n^a = (0,0,1)$, and when the rotation is applied to 
$W_\mu^a$ the inhomogeneous part of the gauge transformation 
contributes in a non-trivial way to $A_\mu$. At late times,
we then find that $\epsilon^{ijk} A_{jk}/2$ coincides with 
$({\bm \nabla}\times {\bf A})^i$, and we can unambiguously
keep track of the total electromagnetic helicity defined as 
${\cal H}(t) = V h(t)$ (see Eq.~(\ref{hnb})).

\begin{figure}
  \includegraphics[width=2.5in,angle=-90]{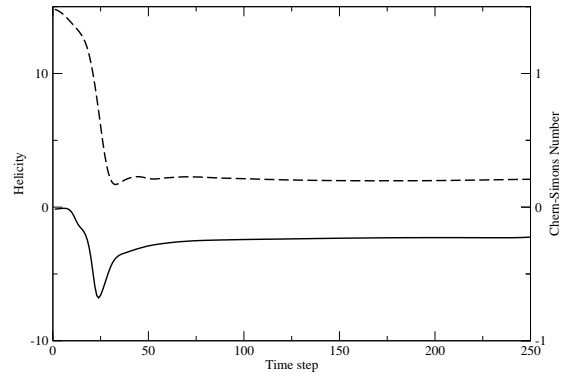}
  \caption{Time evolution of Chern-Simons number (dashed
curve) and electromagnetic helicity (solid curve). The
sign of the Chern-Simons number has been flipped for
greater clarity. 
  }
  \label{ht}
\end{figure}

In Fig.~\ref{ht} we show
the time evolution of the Chern-Simons number and the total 
electromagnetic helicity. We see changes in the Chern-Simons 
number as well as growth in the magnitude of the electromagnetic 
helicity, and then a period of conservation (time step 50
to 250). 
%
While electromagnetic helicity is known to be conserved in a
conducting plasma, its conservation in the present situation
is novel because we are solving vacuum equations and the only
charges in the system are due to the $W^\pm$ fields.

The value of electromagnetic helicity ($\approx 2.5$) is much
less than the value estimated ($\approx 200$) in 
Refs.~\cite{Cornwall:1997ms,Vachaspati:2001nb}. The estimate
in \cite{Vachaspati:2001nb} assumed a certain decay
channel for linked loops of Z-string and it is likely that
the electromagnetic helicity is a function of the precise 
instability through which the sphaleron decays. A more rigorous 
estimate of the final helicity needs to be done statistically, 
taking into account the conditions at the epoch of baryogenesis.

Once electromagnetic field helicity is produced, it
will evolve, and eventually 
get frozen-in in the highly conducting ambient plasma. This is 
the scenario envisaged in Ref.~\cite{Vachaspati:2001nb} 
and, under certain assumptions about the inverse cascade, leads 
to the estimate that the cosmic magnetic field is $\sim 10^{-13}$ G 
at recombination and coherent on a comoving scale of $\sim 0.1$ pc. 
We now argue that this is really an {\it underestimate} of the 
magnetic field strength and typically we can expect a much higher 
field strength.

The essential point is that {\it every} baryon number violating
reaction goes via the sphaleron and produces magnetic fields,
whereas the estimate in \cite{Vachaspati:2001nb} only accounts for the 
magnetic field produced due to the {\it excess} of baryons over 
antibaryons that we see today. To make this clearer, suppose that 
sphalerons decay in some volume to produce $N_b$ baryons while
others decay to produce ${\bar N}_b=N_b-\epsilon$ antibaryons,
where $\epsilon$ is entirely due to fundamental CP violation.
Magnetic fields will be produced in each one of these 
$N_b+{\bar N}_b=2N_b-\epsilon$ sphaleron decays. The baryon 
excess, however, is due to CP violation and is given by 
$N_b-{\bar N}_b=\epsilon$ sphaleron decays. The magnetic field 
produced by just these excess number of reactions is much smaller. 
However, just as the baryons and antibaryons can annihilate, it is 
likely that some of the magnetic fields produced due to baryon and 
antibaryon production will also annihilate. It is hard to estimate 
exactly how much magnetic field survives, but the estimate in 
\cite{Vachaspati:2001nb} is the {\it minimum} value of the magnetic 
field. This value is protected by helicity conservation. 

We will now obtain another estimate with less restrictive, and
more realistic, assumptions about the evolution of the magnetic
fields. Let us suppose that the magnetic fields due to baryon
and antibaryon production do not completely annihilate, leading
to a magnetic field enhancement by a factor $r$. Then
\be
B \sim 10^{-13} r ~ G
\ee
Successful big bang nucleosynthesis (BBN) constrains the energy 
density in the magnetic field so that $B(t_{\rm rec}) < 1$ G. 
We choose the BBN bound as opposed to other bounds since BBN 
places strong constraints on magnetic fields with small coherence 
scale {\it e.g.} see \cite{Kernan:1995bz}. (Other constraints due 
to gravitational wave production may apply to the non-helical 
component of the magnetic field \cite{Caprini:2001nb}.) This 
implies that
\be
r < 10^{13}
\label{rbound}
\ee

Now to connect to particle physics, we realize that $r$ depends on 
the total number of sphaleron processes, which are $N_b+{\bar N}_b$ 
in number. The estimate in \cite{Vachaspati:2001nb} of $10^{-13}$ G 
at recombination assumes that the magnetic field is proportional 
only to the net baryon number $N_b-{\bar N}_b$. Assuming that 
the magnetic field strength gets reduced due to annihilations
in some stochastic way, we expect
\be
r \sim \left ( \frac{N_b+{\bar N}_b}{N_b-{\bar N}_b} \right )^\gamma
\ee
where $\gamma$ is an undetermined exponent which is 1 for no
annihilation and 1/2 for Brownian evolution of the magnetic
field strength.

The baryon excess is purely due to CP violation in the particle 
physics responsible for baryogenesis. For example, in the case of 
the electroweak model~\cite{Riotto:1999yt},
\be
\frac{N_b+{\bar N}_b}{N_b-{\bar N}_b} \sim 10^{20}
\ee
leading to $10^{-3}$ G at recombination for the stochastic case
($\gamma =1/2$) and $10^{7}$ G for the no annihilation 
case ($\gamma =1$). The latter estimate clearly violates
the BBN bound in Eq.~(\ref{rbound}), while the former leads
to $10^{-9}$ G today though on short coherence scales (0.1 pc). 

The argument above is intended to show that it may be possible to
derive important model-independent constraints on particle physics 
from limits on cosmic magnetic fields. Further, if cosmic magnetic 
fields are observed, they can be used to derive detailed information 
about processes at baryogenesis and hence about high energy particle 
physics and CP violation.

We are grateful to Jose Blanco-Pillado, Juan Garc\'{\i}a-Bellido,
Margarita Garc\'{\i}a, Noah Graham, Biagio Lucini
and Jon Urrestilla for discussions. TV thanks the Netherlands Organization 
for Scientific Research (NWO) for a visitor's grant, and the Lorentz 
Institute of Leiden University, and ICTP (Trieste) for hospitality.
This work was supported in part by the U.S. Department of Energy and
NASA at Case Western Reserve University, and by NWO under the VICI
programme.

\end{document}